   \documentclass[apj]{emulateapj}

\usepackage{psfig}

\newcommand{\Msun}{\mbox{$\rm M_{\odot}$}}

\newcommand{\LCDM}{\mbox{$\Lambda$CDM}}

\newcommand{\simlt}{\rlap{$<$}{\lower 1.0ex\hbox{$\sim$}}}
\newcommand{\simgt}{\rlap{$>$}{\lower 1.0ex\hbox{$\sim$}}}

\newcommand{\htwo}{\mbox{H$_2$}} 
\newcommand{\e}{\mbox{e$^-$}}
\newcommand{\h}{\mbox{H}} 
\newcommand{\hplus}{\mbox{H$^+$}}
\newcommand{\htwoplus}{\mbox{H$_2^+$}} 
\newcommand{\hminus}{\mbox{H$^-$}}
\newcommand{\he}{\mbox{He}} 
\newcommand{\heplus}{\mbox{He$^+$}}
\newcommand{\heplusplus}{\mbox{He$^{+2}$}}

\newcommand{\lya}{Ly$\alpha$}

\newcommand{\cii}{[C~II]}
\newcommand{\oi}{[O~I]}
\newcommand{\siii}{[Si~II]}
\newcommand{\feii}{[Fe~II]}

\begin{document}

\title{CRITICAL METALLICITY AND FINE-STRUCTURE EMISSION OF 
PRIMORDIAL GAS \\
ENRICHED BY THE FIRST STARS}

\author{Fernando Santoro \& J. Michael Shull } 

\affil{CASA, Department of Astrophysical and Planetary Sciences, 
University of Colorado, 389-UCB, Boulder, CO 80309; 
fsantoro@casa.colorado.edu, mshull@casa.colorado.edu}



\begin{abstract}

The influence of the first stars on the formation of second-generation
objects at high redshift may be determined, in part, by their metal 
enrichment of surrounding gas.  At a critical metallicity, 
$Z_{\rm crit}$, primordial gas cools more efficiently by fine-structure 
lines of \cii\ (157.74 $\mu$m), \oi\ (63.18~$\mu$m, 145.5~$\mu$m), \siii\ 
(34.8~$\mu$m), and \feii\ (25.99~$\mu$m, 35.35~$\mu$m) than by H~I or \htwo\ 
emission.  This cooling may alter the process of fragmentation into 
smaller units.  We study the time-dependent cooling of primordial gas 
enriched by heavy elements from early massive stars, particularly O, 
Si, and Fe.  We define $Z_{\rm crit}$ as the point when the total 
cooling rate by metals plus \htwo\ equals the adiabatic compressional 
heating. We explore two metallicity scenarios: (1) a single metallicity 
for all heavy elements; (2) individual metallicities 
($Z_{\rm C}$, $Z_{\rm O}$, $Z_{\rm Si}$, $Z_{\rm Fe}$) from theoretical 
supernova yields.  For dense gas ($n \geq 10^3$ cm$^{-3}$) with  
metals in relative solar abundances, fragmentation occurs at 
$Z_{\rm crit} \approx 10^{-3.5} Z_{\odot}$.  However, for lower density 
gas ($n = 1-100$~cm$^{-3}$), particularly in halos enriched in Si, O, 
and Fe, we find $Z_{\rm crit} \approx$ 0.1--1\% $Z_{\odot}$.  The critical 
metallicity approaches a minimum value at high-$n$ set by efficient LTE 
cooling, with thermalized level populations of fine-structure states and 
\htwo\ rotational states ($J = 2$ and $J = 3$).  Primordial clouds of 
$10^8~M_{\odot}$ at 200~K are detectable in redshifted fine-structure 
lines, with far-infrared fluxes between $10^{-22}$ and $10^{-21}$ W~m$^{-2}$.  
For metallicities $Z_{\rm O} \approx 10^{-3}$ and molecular fractions
$f_{\rm H2} \approx 10^{-3}$, the fine-structure emission lines of \oi,  
\siii, and \feii\ could be $10^2$ -- $10^3$ times stronger than the \htwo\ 
rotational lines at 28.22 $\mu$m ($J = 2-0$) and 17.03 $\mu$m ($J = 3-1$). 

\end{abstract}


\section{Introduction}
\label{sec:intro}

The last few years have witnessed an increase in the number of 
studies of first stars \citep{TS00, BL04, CF04}. Implicit 
in these studies is the issue of determining the transition between 
the first (metal-free) stars and the second (metal-enriched) 
generation of stars. This transition is thought to be determined 
primarily by the amount of heavy elements expelled by early stars and 
supernovae (SNe) and by their rate of incorporation into the
surrounding intergalactic medium (IGM) and gaseous halos. 
At temperatures $T \geq 8000$~K, primordial gas cools efficiently 
by the excitation of resonance lines of hydrogen and helium. To cool 
below the ``Ly$\alpha$ barrier" at 8000~K, the gas must form \htwo\ 
\citep{Peeb68,Hira69,LS84}, whose rotational excitation allows the 
gas to cool to lower temperatures ($\sim 10^2$~K) 
and form the first stars \citep{AAZN97}. 

First-generation massive stars end their lives as SNe and eject
newly synthesized heavy elements into interstellar space and out 
into the IGM.  Once these metals are incorporated into neighboring 
clumps and halos, a second generation of stars will form from 
metal-enriched gas.  When this gas includes a sufficient fraction
of metals, it can cool more efficiently via the excitation of
fine-structure levels of key heavy elements (C, O, Si, Fe).  
At this point, the gas is said to have reached ``critical metallicity" 
\citep{BFCL01, BL03}, and the cooling gas is then able to fragment 
into smaller cores and lower-mass stars.

Metal enrichment and the subsequent cooling both depend on the mass 
range of the first stars.  Zero-metallicity stars above $140~M_{\odot}$ 
are predicted to end their lives in pair-instability supernovae or PISN 
\citep{HW02} and eject large amounts of O, Si, and Fe.  Massive stars 
in the range 10--100 $M_{\odot}$ produce smaller amounts of Fe, but 
synthesize considerable mass in $\alpha$-process elements (O, Ne, Si, S, Ar).  
As noted by Tumlinson, Venkatesan, \& Shull (2004) and Qian \&
Wasserburg (2005), the nucleosynthetic signatures of ``Population~III" 
stars provide key diagnostics of the initial mass function (IMF) of the 
first stars and SNe. 
 
The purpose of this paper is to calculate a minimum 
metallicity, $Z_{\rm crit}$, at which the gas reaches the mass 
scale of fragmentation after the death of the first generation of 
stars \citep{SFNO02, BL03, FC04}. Previous work \citep{Omu00}
used semi-analytical models to describe the thermal and chemical 
evolution of a protostellar cloud with a range of metallicities. 
In that study, a large network of chemical reactions was followed 
to describe the non-equilibrium chemistry for four elements 
(H, He, C, O) and 45 species, but no conclusions were drawn about the 
critical metallicity.

\cite{BL03} made an analytic calculation of the critical metallicity,  
using cooling rates from fine-structure lines of [C~II] and [O~I].
However, they did not include other likely coolants 
such as \siii\ or \feii, nor did they follow the 
thermal evolution of the system.  Based on the low [Fe/H] seen in
several EMP (extremely metal poor) halo stars \citep{Chr02}, they 
identified C and O as the elements responsible for the transition 
between Pop~III and Pop~II and derived critical values,  
$\rm{[C/H]}=-3.5$ and $\rm{[O/H]}=-3.05$, for the IGM.  Their study
began with pre-existing clumps of cold, high-density gas.  They 
assumed an initial hydrogen density $n_H=10^4$~cm$^{-3}$ and explored
a range of temperatures, 100~K $<T<$ 200~K, set by \htwo\ cooling. 

\cite{FC04} also explored this transition, but they used different 
IMFs and were concerned with the photon production of the first objects. 
To connect $Z_{\rm crit}$ to IGM reionization, they calculated a 
relationship between metallicity of the IGM and ionizing photons 
from PISN and type-II supernovae (SNeII). In order to reach a 
critical metallicity of $\sim 10^{-3.5}~Z_{\odot}$ for an IMF containing 
very massive stars (VMS), the transition between the two populations 
occurred when $\sim1$~photon per H atom is produced. Approximately 
10 photons per H are produced with a normal top-heavy IMF.

In this paper, we assume that a first generation of massive stars 
formed in a high-redshift halo.  These stars produced a UV background
and distributed the products of nucleosynthesis throughout the halo.   
We assume that all elements with first ionization potentials 
less than 13.6 eV are kept singly ionized (C~II, Si~II, Fe~II, etc).
A similar line of reasoning was followed in \cite{BL03},  
but we introduce three improvements in the astrophysical
calculations.  First, we solve the time-dependent thermal history of 
the gas, cooling isobarically from an initial temperature and density. 
Second, we increase the number of cooling metal lines, both
\cii\ and \oi\ but also \siii\ and \feii.  
Third, we explore deviations from a single metallicity, $Z$,
by defining individual values ($Z_{\rm C}$, $Z_{\rm O}$, $Z_{\rm Si}$, 
$Z_{\rm Fe}$) in accordance with theoretical yields from PISN and VMS.

The paper is structured as follows. We describe our semi-analytical
simulations in \S~\ref{sec:modelm}.  We explain our chemical network
and density evolution as a function of the different cooling functions 
in \S~\ref{sec:chemmodel}. In \S~\ref{sec:metalcool} we give details 
of the metal cooling lines used in the calculations.  In 
\S~\ref{sec:results} we present our results. We discuss and summarize 
our conclusions in \S~\ref{sec:discuss}.

\section{Semi-analytic simulations}
\label{sec:modelm}

We have approached the cooling problem by considering a parcel 
of gas with primordial composition (see Appendix). We give the 
gas an initial temperature $T \approx 100-200$~K and initial 
metallicity and let it cool isobarically, until the gas reaches 
a condition for fragmentation: a rate of cooling faster than
adiabatic heating rate within a local Hubble time.  We adopt an 
initial mass density of hydrogen, $\rho_{\rm vir} \approx 178 \rho_b(z)$,   
as that of a virialized halo of total mass 
$M = (10^6~M_{\odot})M_6$ and baryon mass $(\Omega_b/\Omega_m) M$ 
at fiducial redshift $z \approx 20$. The hydrogen number density
in all forms is given by $n_H = (1-Y) \rho /m_H$, for helium abundance
$Y = 0.244$ by mass and $\rho_{b,0} = 4.21 \times 10^{-31}$
g~cm$^{-3}$, appropriate for the baryon density parameter 
$\Omega_b h^2 = 0.0224 \pm 0.0009$ found by {\it WMAP} \citep{Sper03}.
The virial radius and average hydrogen density of the halo are then,  
\begin{equation}
   R_{\rm vir} = \left[ \frac {3 M/4 \pi} {178 \rho_m} \right]^{1/2} 
     \approx (160~{\rm pc}) M_6^{1/3} \left[ \frac {20}{1+z} \right] \; \\  
\end{equation}
\begin{equation}
   \langle n_{\rm vir}^{(H)} \rangle = \frac {(1-Y) \rho_{\rm vir}} {m_H}     
     \approx (0.27~{\rm cm}^{-3}) \left[ \frac {1+z}{20} \right]^3 \; .
\end{equation} 
Gas within this halo probably contains clumps of higher density,
but this formula provides a characteristic mean value.
This mean density is 178 times larger than the mean density of the IGM,
$\langle n_H \rangle = (1.90 \times 10^{-7}~{\rm cm}^{-3}) h_{70}^{-2} (1+z)^3$.  
We chose a value of $18 \pi^2 \approx 178$ for the density ratio at
virialization, its value for $\Omega = 1$.  For a cosmology with 
$\Omega + \Lambda = 1$ and $\Omega = 0.3$, this ratio is $\sim 340$. 
In general, the total matter density is 
\begin{equation}
\label{eq:ndensity}
  \rho(z)= \left(\Delta_c\over\Omega\right)\rho_{b,0}(1+z)^3 \; ,
\end{equation}
where $\Delta_c(\Omega, \Lambda) \approx 18\pi^2\Omega^{0.45}$, a formula
accurate to 5\% for values of $\Omega$ = 0.15--1.0 \citep{ENF98}.

The condition for fragmentation has been investigated by many 
authors, both analytically and through hydrodynamical numerical
simulations \citep{RO77, BFPR84,YN98,BFCL01,NaU01,Omu01,SFNO02}. 
The basic condition, $t_{\rm cool} \ll t_{\rm ff}$, requires that 
the cooling time be shorter than the free-fall time, the time for 
a perturbation at the surface to reach the center of the clump,
\begin{eqnarray}
   t_{\rm cool} &=& \frac {3n_t kT} {2 {\cal L}(n, T)} \nonumber \\ 
   t_{\rm ff}   &=& \left( \frac {3\pi} {32G\rho} \right)^{1/2} \; . 
\end{eqnarray}
Here, $n_t$ is the total particle number density and
${\cal L}(n, T)$ is the cooling rate per unit volume
(erg~cm$^{-3}$~s$^{-1}$). In this cooling rate, we include 
[C~II], [Si~II], [O~I] and [Fe~II] fine-structure and forbidden
lines, as well as cooling by H~I and \htwo.  These four heavy
elements (C, O, Si, Fe) are products of $\alpha$-burning
whose dominant interstellar ions have efficient cooling from low-lying 
fine-structure levels.  Other $\alpha$-process species such 
as Ne~I, Mg~II, S~II, Ar~I, and Ca~II
lack fine structure states in their ground configurations. 
We refer the reader to \S~\ref{sec:atomicdata} and 
\S~\ref{sec:metalcool} for a complete description of the 
atomic data and metal cooling rates.

As the cloud contracts, with sufficient cooling for framentation, 
the temperature drops and the density increases. The cloud sizes at
fragmentation are characterized 
by the Jeans length scale and Jeans mass, $M_J$.  Fragmentation slows when 
the gas becomes optically thick, but also when the gas density exceeds 
the critical densities for collisional de-excitation of the dominant 
coolants.  This defines a minimum mass, which we use as the condition for 
fragmentation.  This condition depends on the choice of the cooling
function. \cite{BL03} studied the formation of low-mass stars from gas
that has been ``polluted" by trace amounts of carbon and oxygen. To
derive the critical metal abundances for fragmentation, they
looked for the point at which the $\it{metal}$-$\it{only}$ cooling 
rate, ${\cal L}_{\rm CII,OI}$, equals the adiabatic compressional 
heating rate, $\Gamma_{\rm ad}$.  

The gas fragments when ${\cal L} = \Gamma_{\rm ad}$, where 
$\Gamma_{\rm ad} = -P (d~{\rm ln} \rho /dt) = 3P (dR/dt)/R$ 
is the adiabatic compression heating per unit volume 
(erg~cm$^{-3}$~s$^{-1}$) arising from free-fall collapse 
of the gas under gravity, 
\begin{equation}
   \Gamma_{\rm ad} = (24\pi)^{1/2}~[1-R/R_{0}]^{1/2}(n_{\rm{t}}kT)
   \sqrt{G\rho}  \; .
\end{equation}
Here, $\rho = 1.323 n_H m_H$ is the total mass density, where
$n_H$ is the hydrogen number density and $n_t = 1.0807 n_H$ 
is the total particle number density (mostly H and He, since the
electron and molecular fractions are quite small). As before, these 
values assume helium mass fraction $Y = 0.244$ (He/H = 0.0807 by 
number). 

The initial and current cloud radii are $R_0$ and $R$, and the 
coefficient $(24\pi)^{1/2}~[1-R/R_{0}]^{1/2}$~ varies from 
1.22 to 2.74 for clouds with radius $(R/R_0)$ =0.99--0.90. We use 
the value corresponding to a $10\%$ change in radius, for a maximum 
coefficient (2.74), and we write the adiabatic heating rate 
(erg~cm$^{-3}$~s$^{-1}$) as, 
\begin{equation}
   \Gamma_{\rm ad} = (4.59 \times 10^{-30})
   \left( \frac {T}{200~{\rm K}} \right)
   \left( \frac {n_{\rm t}} {0.3~{\rm cm}^{-3}} \right)^{3/2} \; . 
\label{eq:adheat}
\end{equation}

The different terms that form part of the cooling function used in
these definitions are explained in the following section.
We use the fragmentation criterion ${\cal L} = \Gamma_{\rm ad}$  
and explore cases in which adiabatic heating is compared to 
metal-line cooling only, as well as to the total cooling.  In practice,
both methods give similar results.

\subsection{Chemistry network and thermal evolution}
\label{sec:chemmodel}

Throughout this paper we assume a \LCDM ~cosmology with
$\Omega_0 = 0.30$, $\Omega_{\Lambda} = 0.70$, $\Omega_{b0} = 0.0457$, 
$h = 0.7$, and $\sigma_8 = 0.90$, the root-mean-square density 
dispersion within spheres of radius $8h^{-1}$Mpc. 
As initial conditions, we adopt residual electron and molecular fractions
$n_{\rm e,res}/n_{\rm H} = 5 \times 10^{-4}$~\citep{Seag00} 
and $n_{\rm {H_2,res}}/n_{\rm H} =2 \times10^{-4}$~\citep{LSD02}.
These fractions are ``frozen out" from expansion following the
recombination epoch.  
We model the gas as $75.6\%$ H and $24.4\%$ He by mass and 
include traces amount of metals as coolants. We follow the 
non-equilibrium chemistry of nine species, \htwo, \h, \hplus, 
\htwoplus, \hminus, \he, \heplus, \heplusplus, and \e\   
\citep{HST02}. The reaction rates and chemical processes can be 
found in the Appendix. 

We follow the evolution of the gas density at constant pressure, $P$. 
From the energy equation, we find, 
\begin{equation}
  \frac{d n_t}{dt} = -\frac{n_t~{\cal L}} {P} \; , 
\label{eq:enereq}
\end{equation}
where $n_{\rm{\small{t}}}$~is the total particle number density and
${\cal L}$ is the cooling rate per volume. From the ideal gas relation,
we obtain the temperature, $T = P/n_t k$, and define 
\begin{equation}
   {\cal L} = {\cal L}_{\rm metals} + {\cal L}_{\rm{H_2}} 
      + {\cal L}_{\rm{H}} + {\cal L}_{\rm{He^+}} \; , 
\label{eq:coolfuncs}
\end{equation}
where the individual terms denote the cooling rates from metals, \htwo, 
H$^{\circ}$, and He$^+$.  

To find the molecular hydrogen cooling function, we solve for the
statistical equilibrium of the first six rotational levels ($J$ = 0--5). 
We use the radiative transition rates for the \htwo\ rotational
quadrupole lines from Wolniewicz, Simbotin, \& Dalgarno (1998): 
$A(J = 2\rightarrow 0) = 2.94 \times 10^{-11}$~s$^{-1}$,
$A(J = 3\rightarrow 1) = 4.76 \times 10^{-10}$~s$^{-1}$, 
$A(J = 4\rightarrow 2) = 2.75 \times 10^{-9}$~s$^{-1}$, and 
$A(J = 5\rightarrow 3) = 9.83 \times 10^{-9}$~s$^{-1}$.  
We assume separate populations of ortho-\htwo\ ($J$ = 1, 3, 5) and
para-\htwo\ ($J$ = 0, 2, 4) with the usual 3:1 ortho/para ratio from 
nuclear spin multiplicities.  The rate coefficients for H$^{\circ}$ 
and He$^{\circ}$ collisional excitation of these levels were taken 
from \cite{FBDL97}, who calculated new rates of (non-reactive) rotational
excitation over the range 10~K $< T <$ 1000~K, using exact vibrational
wavefunctions and a more reliable H$_3$ potential energy surface 
\citep{Boothroyd96} for the interaction potential. 
Their H$^{\circ}$--\htwo\ rates are several times larger than those 
\citep{LPF99} commonly used in other studies.   
We include He--\htwo\ collisions, whose rate coefficients are 30--40 
times larger than those of H$^{\circ}$. We scale the He rates to 
H rates of \cite{FBDL97}, using ratios of He/H rate coefficients 
computed by \cite{LPF99}.

For the complete set of collisional excitation rates and other
less important processes together with their analytical fits see
\cite{HST02}.  The atomic data for the metal fine-structure
and forbidden lines and the cooling rates, ${\cal L}_{\rm metals}$, 
are explained in the following sub-sections.
Table \ref{tab:mabundances} shows the reference solar
metallicities used in this work, taken from \cite{Mo03} for C, O, Si, 
and Fe. The table also lists the critical number densities,  
$n_{\rm cr} = A_{21}/\gamma_{21}$, for H$^{\circ}$ de-excitation of the 
fine-structure lines, where the level populations approach Boltzmann (LTE) 
populations.  Here, $A_{21}$ and $\gamma_{21}$ are the radiative 
decay rate and H$^{\circ}$-impact de-excitation rate coefficient
respectively.  For $n > n_{\rm cr}$, the volume rate of line cooling 
scales as ${\cal L} \propto n Z_{\rm crit}$, rather than $n^2 Z_{\rm crit}$ 
at low density.  Because $\Gamma_{\rm ad} \propto T n^{3/2}$, 
the critical metallicity $Z_{\rm crit} \propto T n^{3/2}/{\cal L}$  
exhibits an approximately parabolic shape around its minimum at 
$n \approx n_{\rm cr}$ for a particular coolant. For a given heavy element, 
$Z_{\rm crit}$ scales as $n^{-1/2}$ for $n < n_{\rm cr}$, and 
as $n^{1/2}$ for $n > n_{\rm cr}$.  The most efficient cooling occurs at
$n \approx n_{\rm cr}$ for the fine-structure lines of a given element.  \\

\begin{table}
\begin{center}
\caption{Metals and Transitions Used in Calculations\tablenotemark{a}}

\label{tab:mabundances}

\begin{tabular}{lccccc} 
\hline\hline \\
 Species   &  $\lambda(\mu{\rm m}$) & $(A_i/A_H)_{\odot}$ & 
   $(A_i/A_{\rm Fe})_{\odot}$ & $n_{cr}$(cm$^{-3}$)\tablenotemark{b} \\ 
\hline \\[2pt] 
C~II  & $157.74$       & $2.45(-4)$ & $7.75$ & $2.86(3)$\\ 
O~I   & $63.18$        & $4.90(-4)$ & $15.5$ & $6.08(5)$\\ 
Si~II & $34.8$         & $3.63(-5)$ & $1.15$ & $2.75(5)$\\ 
Fe~II & $25.99,35.35$  & $3.16(-5)$ & $1.00$ & $2.24(6), 1.46(6)$ \\ 
\hline  
\tablenotetext{a}{Notation: $2.45(-4)$ denotes $2.45\times10^{-4}$, 
  $\lambda$ is the wavelength of the fine-structure transition, and 
  $(A_i/A_H)_{\odot}$ is the solar abundance of element (i) relative 
  to hydrogen (Morton 2003).}
\tablenotetext{b}{$n_{cr}$~is the critical density of the transition, 
  at which H$^{\circ}$-impact collisional de-excitation ($T = 200$~K) equals 
  radiative decay. }
\end{tabular}
\end{center}
\end{table}

\subsection{Atomic Data for Metal Lines} 
\label{sec:atomicdata}
 
\subsubsection{C~II fine-structure line}

The ground state ($2p$) configuration of C~II is treated as a two-level 
atom, with a fine-structure radiative transition from 
$(2p)~[^2P_{3/2} \rightarrow ^2P_{1/2}]$. The atomic parameters were
taken from \cite{HMc89}. This transition corresponds to $157.74~\mu$m 
with excitation energy $E_{21} = 1.259 \times 10^{-14}$ erg and 
excitation temperature, $T_{\rm exc} = E_{21}/k = 91.2$~K. The 
radiative transition rate is $A_{21}=2.4\times10^{-6}$~s$^{-1}$, and 
the H$^{\circ}$-impact de-excitation rate coefficient is:
\begin{equation}
   \gamma^{\rm{\small{H}}}_{21} = 
      (8.0 \times 10^{-10}~{\rm cm}^3~{\rm s}^{-1}) T_{100}^{0.07} \; , 
\end{equation}
where $T_{100} = (T/100~{\rm K})$. 
The critical density for H$^{\circ}$ de-excitation is
$n_{\rm cr} = A_{21}/\gamma_{21} = (3000~{\rm cm}^{-3}) T_{100}^{-0.07}$.
The cooling rate is 
\begin{equation}
  {\cal L}_{\rm CII} =\frac {\gamma^{\rm{\small{H}}}_{12} A_{21}
  E_{21} Z_{\rm C} A_C n_{\rm{\small{H}}}}
      {\gamma^{\rm{\small{H}}}_{21} + 
     \gamma^{\rm{\small{H}}}_{12} + A_{21}/n_{\rm{\small{H}}}} \; , 
\label{eq:cIIcooling}
\end{equation}
where $Z_C$ is the carbon metallicity and $A_C$ is the 
solar carbon abundance relative to hydrogen. By detailed balance,
we have,
\begin{equation}
   \gamma^{\rm{\small{H}}}_{12} = (g_2/g_1) \gamma^{\rm{\small{H}}}_{21} 
   \exp (- E_{21}/kT) \; ,  
\end{equation}
where $g_2=4$ and $g_1=2$ are statistical weights.

\subsubsection{O~I fine-structure and forbidden lines}

The ground-state ($2p^4$) configuration of O~I is a 5-level multiplet,
with transitions between fine structure levels arising from
($^3$P$_2$,$^3$P$_1$,$^3$P$_0$,$^1$D$_2$, $^1$S$_0$). 
We label these levels, from lowest to highest, as 1--5;
their excitation temperatures from the ground ($^3$P$_2$) state
are $T_{\rm exc}$ = 227.7~K ($^3$P$_1$) and 326.7~K ($^3$P$_0$).  
For the excited terms, $T_{\rm exc}$ = 22,523~K ($^1$D$_2$) and 
48,661~K ($^1$S$_0$).   
The fine-structure transitions within the $^3$P term lie at
far-infrared wavelenegths: $63.18~\mu$m ($J = 1 \rightarrow 2$) 
and $145.53~\mu$m ($J = 0 \rightarrow 1$). 
The optical forbidden lines lie at 6391.5~\AA, 6363.8~\AA, and 
6300.3~\AA\ (from $^1$D to $^3$P) and at 2972.3~\AA, 2958.4~\AA, 
and 5577.4~\AA\ from the $^1$S term.

The radiative transition probabilities were taken from tabulations
in \cite{O89}:
  $A_{21}=8.9\times10^{-5}$~s$^{-1}$,
  $A_{31}=1.0\times10^{-10}$~s$^{-1}$,
  $A_{32}=1.7\times10^{-5}$~s$^{-1}$,
  $A_{41}=6.3\times10^{-3}$~s$^{-1}$,
  $A_{42}=2.1\times10^{-3}$~s$^{-1}$,
  $A_{43}=7.3\times10^{-7}$~s$^{-1}$,
  $A_{51}=2.9\times10^{-4}$~s$^{-1}$,
  $A_{52}=7.3\times10^{-2}$~s$^{-1}$,
  $A_{54}=1.2$~s$^{-1}$.
The H$^{\circ}$ collisional de-excitation rate coefficients
(cm$^3$~s$^{-1}$) were taken from \cite{HMc89}:
$\gamma^{\small{H}}_{21}=9.2\times10^{-11} T_{100}^{0.67}$,
$\gamma^{\small{H}}_{31}=4.3\times10^{-11} T_{100}^{0.80}$,
$\gamma^{\small{H}}_{32}=1.1\times10^{-10} T_{100}^{0.44}$.
The remainder of the rate coefficients are less well determined and
are estimated to be $10^{-12}$~cm$^3$~s$^{-1}$.
The critical density of the first excited state, $^3$P$_1$ by 
H$^{\circ}$ de-excitation is
$n_{\rm cr} = (9.67 \times 10^5~{\rm cm}^{-3}) T_{100}^{-0.67}$.
The radiative cooling rate from [O~I] was computed from the equilibrium 
populations, multiplied by their radiative decay 
rates and line energies.

\subsubsection{Si~II fine-structure line}

For the Si~II ($3p$) ground-state configuration, we also   
provide a two-level treatment, with a fine-structure 
radiative transition from $(3p)~[^2P_{3/2} \rightarrow ^2P_{1/2}]$.  
This transition corresponds to 34.8~$\mu$m with 
$E_{21} = 5.71 \times 10^{-14}$ erg and $T_{\rm exc} = 410$~K. 
The atomic parameters were taken from \cite{HMc89}. The spontaneous 
transition rate is $A_{21} = 2.1\times10^{-4}$~s$^{-1}$, 
and the H$^{\circ}$-impact de-excitation rate coefficient is:
\begin{equation}
  \gamma^{\rm{\small{H}}}_{21} = (8.0 \times10^{-10}~{\rm cm}^3~{\rm s}^{-1})
   T_{100}^{-0.07}  \; . 
\end{equation}
The critical density for H$^{\circ}$ de-excitation is
$n_{\rm cr} = (2.62 \times 10^5~{\rm cm}^{-3}) T_{100}^{0.07}$.
As in equation \ref{eq:cIIcooling}, the cooling rate is:
\begin{equation}
 {\cal L}_{\rm{SiII}}=\frac{\gamma^{\rm{\small{H}}}_{12} A_{21} \, 
  E_{21} \,Z_{\rm Si} \, A_{\rm Si} \,  
  n_{\rm{\small{H}}}}{\gamma^{\rm{\small{H}}}_{21} + 
   \gamma^{\rm{\small{H}}}_{12} + A_{21}/n_{\rm{\small{H}}}}  \; . 
\label{eq:SiIIcooling}
\end{equation}

\subsubsection{Fe~II fine-structure lines}

The ground-state ($^6$D) configuration of Fe~II is modeled as a 
5-level atom, with fine-structure levels of total angular momentum 
$J$ = 9/2, 7/2, 5/2, 3/2, and 1/2 (labeled 1--5 respectively). The 
fine-structure level energies were taken from the NBS tables 
\citep{Moore52}, and the radiative transition rates were adopted
from calculations of the ``Iron Project" \citep{QDZ96}. 
The fine-structure levels 2--5 have excitation temperatures 
$T_{\rm exc}$ = 553.6~K, 960.6~K, 1241.1~K, and 1405.7~K above ground.  
The primary radiative transitions occur at mid-infrared wavelengths:
25.99~$\mu$m ($J = 7/2 \rightarrow 9/2$), 
$35.35~\mu$m ($J = 5/2 \rightarrow 7/2$),
$51.28~\mu$m ($J = 3/2 \rightarrow 5/2$), and
$87.41~\mu$m ($J = 1/2 \rightarrow 3/2$).
The spontaneous transition rates are:
  $A_{21}=2.13\times10^{-3}$~s$^{-1}$,
  $A_{32}=1.57\times10^{-3}$~s$^{-1}$,
  $A_{43}=7.18\times10^{-4}$~s$^{-1}$,
  $A_{54}=1.88\times10^{-4}$~s$^{-1}$, taken from \cite{QDZ96}, and
  $A_{31}=1.50\times10^{-9}$~s$^{-1}$, taken from \cite{HMc89}.

The H$^{\circ}$-impact de-excitation rate coefficients were taken  
from \cite{HMc89}, with no assumed temperature dependence. These 
rates (in units cm$^3$ s$^{-1}$) are:
$\gamma^{\small{H}}_{21}=9.5\times10^{-10}$, 
$\gamma^{\small{H}}_{31}=5.7\times10^{-10}$ and 
$\gamma^{\small{H}}_{32}=4.7\times10^{-10}$.
In the absence of information on higher-$J$ levels in the $^6$D 
multiplet, we have adopted de-excitation rates of 
$5 \times 10^{-10}$ cm$^3$~s$^{-1}$, constant with temperature.  
The critical density of the first excited state, $^6$D$_{7/2}$,
resulting from H$^{\circ}$ de-excitation is 
$n_{\rm cr} = 2.24 \times 10^6~{\rm cm}^{-3}$. 
We computed the radiative cooling rate from \feii\ from the 
statistical equilibrium populations of the excited fine-structure 
levels, multiplied by their radiative decay rates and line energies.

\subsection{Metal cooling}
\label{sec:metalcool}

At $T < 8000$~K, excitation of \lya\ is no longer an efficient cooling
process.  Similarly, at $T < 60$~K, cooling by  pure rotational lines
from \htwo\ becomes ineffective.  At very low temperatures
($T  < 60$~K, HD cooling can be effective, despite the small
primordial abundance, (D/H)$_{\rm prim} \approx 26$ ppm, owing to
its weak dipole transitions \citep{LS84,LSD02}.
At $T < 1000$~K, in the absence of external photoionizing sources,
the hydrogen gas is completly neutral, and low-energy fine-structure
transitions of \cii, \oi, \siii, and \feii, excited by H$^{\circ}$
collisions, become the most important coolants.  (We assume that
C, Si, and Fe are photoionized by ambient UV radiation from newly
formed hot stars.)  These fine-structure transitions can also be
excited by electrons, once the electron fraction, $x_e$,
rises above about $10^{-3}$.  At redshifts, $z < 20$, the residual
electron fraction is $x_e \approx 10^{-3.3}$, and the electrons
contributed by photoionization of trace metals (C$^+$, Si$^+$, S$^+$,
Fe$^+$) are negligible. As the cloud cools and condenses, $x_e$
declines owing to recombination of the trace H$^+$.

\begin{figure}
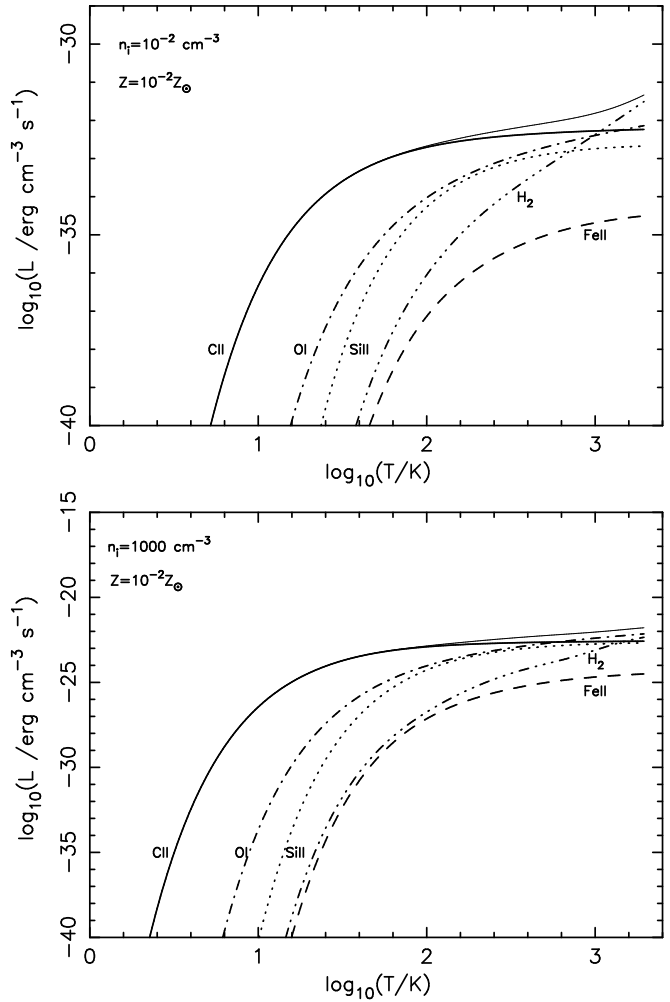

\psfig{width=8.7cm,angle=270,file=Fig1a.ps}
\psfig{width=8.7cm,angle=270,file=Fig1b.ps}
\caption{Total cooling rate at $Z = 0.01~Z_{\odot}$ (thin full line) and
  individual rates for \cii\ (thick full line), \htwo\ (dash-three-dotted
  line), \oi\ (dash-dotted line), \siii\ (dotted line) and \feii\
  (dashed line).  The gas is assumed to have total number densities 
  $n_{\rm{t}}=10^{-2}$ cm$^{-3}$ (top panel) and $10^3$ cm$^{-3}$ 
  (bottom panel). }
\label{fig:coolhighZ}
\end{figure}

\begin{figure}
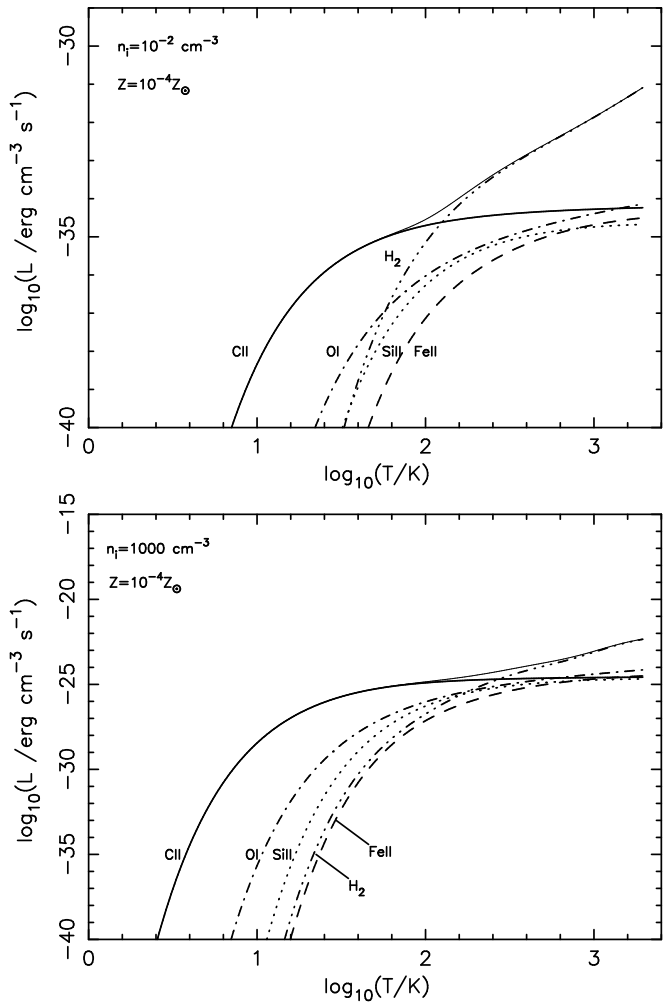

\psfig{width=8.7cm,angle=270,file=Fig2a.ps}
\psfig{width=8.7cm,angle=270,file=Fig2b.ps}
\caption{Cooling rates, as in Fig. 1, for gas with metallicity
  $Z = 10^{-4}Z_{\odot}$ and densities $n_{\rm{t}}=10^{-2}$ cm$^{-3}$
  (top panel) and $10^3$ cm$^{-3}$ (bottom panel). Metals have abundances
  in solar ratios.  }
\label{fig:coollowZ}
\end{figure}

In Figures \ref{fig:coolhighZ} and \ref{fig:coollowZ} we plot the
four metal cooling functions together with ${\cal L}_{\rm H2}$ at
metallicities of $10^{-2} Z_{\odot}$ and $10^{-4} Z_{\odot}$, respectively.
In both figures, we show the total cooling function, as well as
individual rates from \htwo, \cii, \oi, \siii, and \feii.
The low-density cooling rates are for gas with total number
density $n_t = 10^{-2}~{\rm cm}^{-3}$.
At high densities ($n_H \geq 10^3$ cm$^{-3}$) and low metallicities
($Z \leq 10^{-2}~Z_{\odot}$), cooling by H$^{\circ}$ excitation of
\htwo\ becomes important at $T < 200-400$~K compared to the metal lines. \\

\subsection{Analytic Estimates of Critical Metallicities}

The criterion for fragmentation, used to derive $Z_{\rm crit}$,
comes from equating heating and cooling functions of the form,
$\Gamma_{\rm ad} = a T n^{3/2}$ and ${\cal L}(n,T)$.     
From equation (5), we have $a = 2.74 k (G \mu)^{1/2}$,
where $\mu = 1.323 m_H$.  Above the critical density,
when the fine-structure levels reach LTE (Boltzmann) populations, 
the LTE cooling function can be written ${\cal L} = n Z f(T)$, 
where $f(T)$ can be derived from the LTE excited populations times 
the radiative transition rates ($A_{21}$) and line energies 
($E_{21}$). Thus, the critical metallicity is  
\begin{equation}
   Z_{\rm crit}/Z_{\odot} = \left[ \frac {a \, T} {f(T)} \right] n^{1/2} \, 
\end{equation}
The LTE cooling function, $f(T)$, can easily be evaluated for 
a two-level atom of element (i) of solar abundance $A_i$
relative to hydrogen: 
\begin{equation}
  f(T) = \frac {A_i \, A^{(i)}_{21} \, E^{(i)}_{21} } 
        { [ 1 + (g_1/g_2) \exp (E^{(i)}_{21}/kT) ] }  \; ,   
\end{equation}
where $g_1$ and $g_2$ are the statistical weights of the lower
and upper states, respectively, of the two-level transition.   
From the above, we arrive at a convenient analytic expression
for the critical metallicity for a given atom and fine-structure
coolant, that depends only on its reference abundance
and the atomic parameters of the transition:
\begin{equation}
  Z_{\rm crit}/Z_{\odot} = \left[ \frac {2.74 k (G \mu)^{1/2}}  
            {A_i A^{(i)}_{21} E^{i}_{21} } \right] 
          \left[ 1 + \frac {g_1}{g_2} \exp (E^{(i)}_{21}/kT) \right]  
       n^{1/2} T 
\end{equation}
We can evaluate this expression for the two-level species \cii\ and \siii\ 
at $n \geq n_{\rm cr}$ and $T = (200~{\rm K})T_{200}$, where efficient 
fine-structure LTE cooling begins:   
\begin{eqnarray} 
Z^{\rm (CII)}_{\rm crit}  &=& (3.74 \times 10^{-4} Z_{\odot}) 
    \left( \frac {n}{n_{\rm cr}} \right)^{1/2} \\ 
Z^{\rm (SiII)}_{\rm crit} &=& (8.34 \times 10^{-5} Z_{\odot}) 
    \left( \frac {n}{n_{\rm cr}} \right)^{1/2} \; .  
\end{eqnarray} 
For \oi\ and \feii, with 3-level and 5-level fine-structure states,
respectively, we can apply the 2-level formalism to the lowest two states,
to find approximate values,
\begin{eqnarray}
Z^{\rm (OI)}_{\rm crit}   &\approx& (1.02 \times 10^{-4} Z_{\odot}) 
   \left( \frac {n}{n_{\rm cr}} \right)^{1/2} \\
Z^{\rm (FeII)}_{\rm crit} &\approx& (1.76 \times 10^{-4} Z_{\odot}) 
    \left( \frac {n}{n_{\rm cr}} \right)^{1/2}  \; .
\end{eqnarray}
The critical densities for H$^{\circ}$ de-excitation are: 
$n_{\rm cr}({\rm CII})  =  (2860~{\rm cm}^{-3}) T_{200}^{-0.07}$, 
$n_{\rm cr}({\rm OI})   =   (6.08 \times 10^5~{\rm cm}^{-3}) T_{200}^{-0.67}$, 
$n_{\rm cr}({\rm SiII}) =   (2.75 \times 10^5~{\rm cm}^{-3}) T_{200}^{0.07}$, and 
$n_{\rm cr}({\rm FeII}) =   (2.24 \times 10^6~{\rm cm}^{-3})$.
These formulae provide reference metallicities, but they are no substitute
for the exact calculations described below.  For instance, they do not give 
the temperature dependence implicit in equation (16), arising from the Boltzmann 
factor for excitation of fine-structure states.  However, for LTE conditions
at high density, these formulae agree fairly well with the numerical computations.  
Small differences occur for the multi-level \feii\ and \oi\ lines,
since we used a 2-level approximation that becomes inaccurate at higher
temperatures by omitting higher states.  

\section{Results}
\label{sec:results}

In \S~\ref{sec:singlez} we show results from simulations with a 
single metallicity, $Z$, adopting the same relative metal abundances 
for each of the four metals considered.   
In \S~\ref{sec:multiplez} we include individual metallicities
for C, O, Si, and Fe, using metal yields from PISN calculated
by \cite{TVS04}. In our formulation, each metal appears with a
different abundance relative to the other three elements,
depending on the mass range of progenitors.  Our analysis of 
$Z_{\rm crit}$ follows the fragmentation criterion (\S~2), in
which we compare the adiabatic heating with cooling rates from 
metals plus \htwo.  We also present the final properties of the 
fragmented gas, such as temperature and Jeans mass, 
for a variety of metallicities and densities.

\subsection{Single-metallicity simulations}
\label{sec:singlez}

\begin{figure}
\psfig{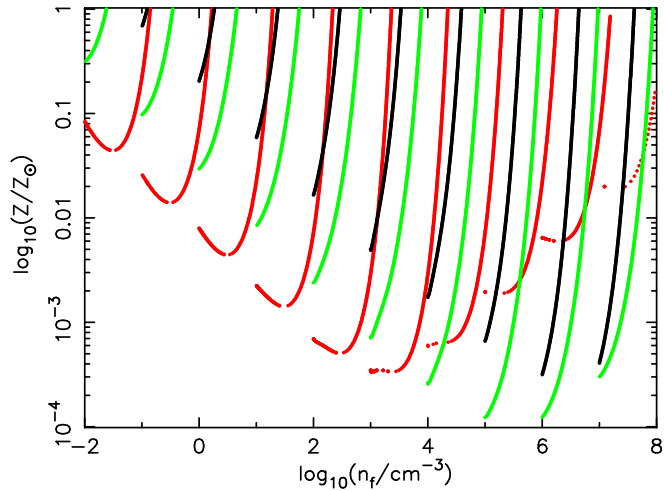}
\caption{Critical metallicity as a function of final number
  density. Each parabola shows the density evolution for a gas for 
  10 different {\it initial} number densities. From left to right: 
  $10^{-2}~{\rm cm}^{-3}$ to $10^{7}$~cm$^{-3}$~in order-of-magnitude 
  steps. Red corresponds to gas composed only of hydrogen plus trace 
  C~II, green shows Si~II and O~I, and black shows Fe~II. } 
\label{fig:paracoolmcool}
\end{figure}

First, we assume that all metals in the gas appear with the same 
relative abundance, characterized by a single metallicity ($Z$). 
Figure \ref{fig:paracoolmcool} synthesizes the results of many 
models of cloud evolution, for different initial densities, with an
initial temperature of 200~K. 
The elements are color-coded in groups (C-alone, O plus Si, Fe-alone)
reflecting their probable origin from progenitor mass ranges and
nucleosynthetic pathways.
There are three set of curves corresponding to three choices of the
cooling function.  In red, we use only metal cooling by \cii.
In green, we use the combined metal cooling function of \siii\ and \oi,
and black corresponds to only \feii.
We have combined the results for \siii\ and \oi, since these elements
come from similar massive stars, and because this presentation makes
the results less cluttered. The \oi\ 63~$\mu$m line dominates the
cooling when only those two elements are considered.

\begin{figure}
\psfig{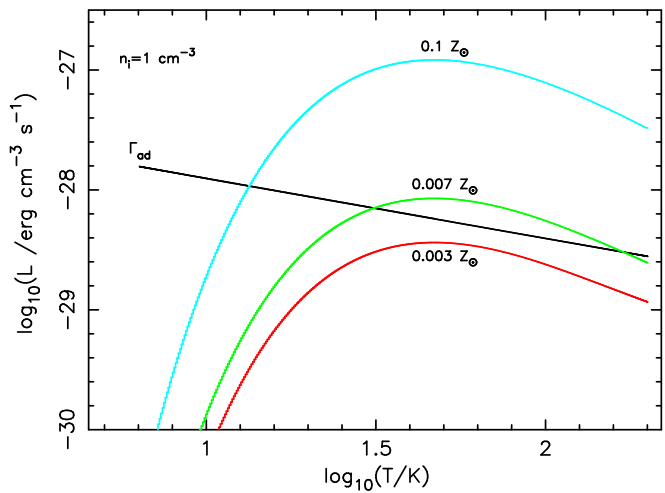}
\caption{Radiative cooling rates and equilibria for gas with 
$n_i = 1$ cm$^{-3}$ and three metallicities. The adiabatic heating 
rate is also shown. Note the double solutions at log~$T$ = 1.5 and 
2.2 for $Z = 0.007~Z_{\odot}$. These two thermal equilibria are 
reflected in the parabolas in Figure 3. There are no equilibria at 
$Z = 0.003~Z_{\odot}$, so that fragmentation does not occur.  }
\label{fig:coolingequilib}
\end{figure}

Each set of three curves (red, green, black) in Figure 
\ref{fig:paracoolmcool} corresponds to the locus of fragmentation points 
in metallicity $Z$ and final density $n_f$.  The code starts with an 
initial density, corresponding to the point farthest to the left of 
each curve.  Each point on the 10 sets of curves is found by running many 
simulations for various initial densities and metallicities, until 
they satisfy the fragmentation criterion (${\cal L} = \Gamma_{\rm ad}$) 
at $Z_{\rm crit}$ and density $n_f$.  The double-valuedness in many
of the parabolic curves is explained by Figure \ref{fig:coolingequilib},
which shows the cooling rates and equilibria for gas with 
$n_i = 1$ cm$^{-3}$ and three metallicities.

Table \ref{tab:lowhighdcritm} summarizes the results of Figure
\ref{fig:paracoolmcool} by choosing three particular values of 
temperature and density relevant for this study.  The first column 
shows the temperature and density of the gas. For the low-density 
case, we choose the set of three curves corresponding to an
initial density of 0.1 cm$^{-3}$, while the high-density case 
starts at $10^4$~cm$^{-3}$. We begin at $T_i = 200$~K, and after 
the gas cools down, we choose two more temperatures, 150~K and 100~K 
and their corresponding densities. The next four columns
show the metallicity at those temperatures and densities during 
fragmentation. In the fourth column, all metal lines are included in 
the cooling function. The last column shows the Jeans mass, 
$M_J \propto T^{3/2}/\rho^{1/2}$, at fragmentation,  
\begin{equation}
  M_J = (700~M_{\odot}) \left( \frac {T_{\rm frag}} {200~{\rm K}} \right)^{3/2}
  \left( \frac {n_H} {10^4~{\rm cm}^{-3}} \right)^{-1/2} \; , 
\label{eq:jeansmass}
\end{equation}
taken from \cite{CB03}, where $T_{\rm{frag}}$~ is the temperature at
fragmentation and $n_H$ is the hydrogen number density.

\begin{table}
\begin{center}
\caption{Critical Metallicities\tablenotemark{a} (log~$Z_{\rm crit}$)}

\label{tab:lowhighdcritm}

\begin{tabular}{cccccc}
\hline\hline \\
($T,~n$) & C~II\tablenotemark{b} & (O~I,~Si~II) &
      Fe~II & All Metals & $M_J$\tablenotemark{c} \\
\hline \\
Low $n$ & & & & &\\
\hline
 200, 1.0(-1) & $-1.59$  & $-1.01$ & $-0.16$ & $-1.70$ &$2.1(5)$\\
 150, 1.3(-1) & $-1.69$  & $-0.92$ &  ...    & $-1.77$ &$1.2(5)$\\
 100, 2.0(-1) & $-1.80$  & $-0.67$ &  ...    & $-1.83$ &$5.7(4)$\\
\hline \\
High $n$ & & & & & \\
\hline
 200, 1.0(4) & $-3.22$  & $-3.58$ & $-2.75$ & $-3.78$ &$698$\\
 150, 1.3(4) & $-3.20$  & $-3.44$ & $-2.46$ & $-3.66$ &$353$\\
 100, 2.0(4) & $-3.19$  & $-3.16$ & $-1.86$ & $-3.49$ &$135$\\
\hline
\tablenotetext{a}{Minimum values of log~$Z_{\rm crit}$ together with
  {\it final} values of density $n$ (in cm$^{-3}$) and temperature 
  $T$~(K) at the point of fragmentation.  For \feii, we found no
  sensible low-$Z$ solutions at $T = 100$~K and 150~K, since its
  first cooling level lies at $T_{\rm exc} = 554$~K. }
\tablenotetext{b}{Columns 2--5 give the critical metallicity,
  in units of solar metallicity, corresponding to a choice of metals 
  in the cooling function. Column C~II means that only \cii\ lines are 
  included; Column labeled ``All Metals" includes C, O, Si, Fe. }
\tablenotetext{c}{Jeans mass $M_J$ in $M_{\odot}$. Same format as in
  Table \ref{tab:mabundances}, where 1.0(4) denotes $1.0 \times 10^4$.}
\end{tabular}
\end{center}
\end{table}

\begin{figure}
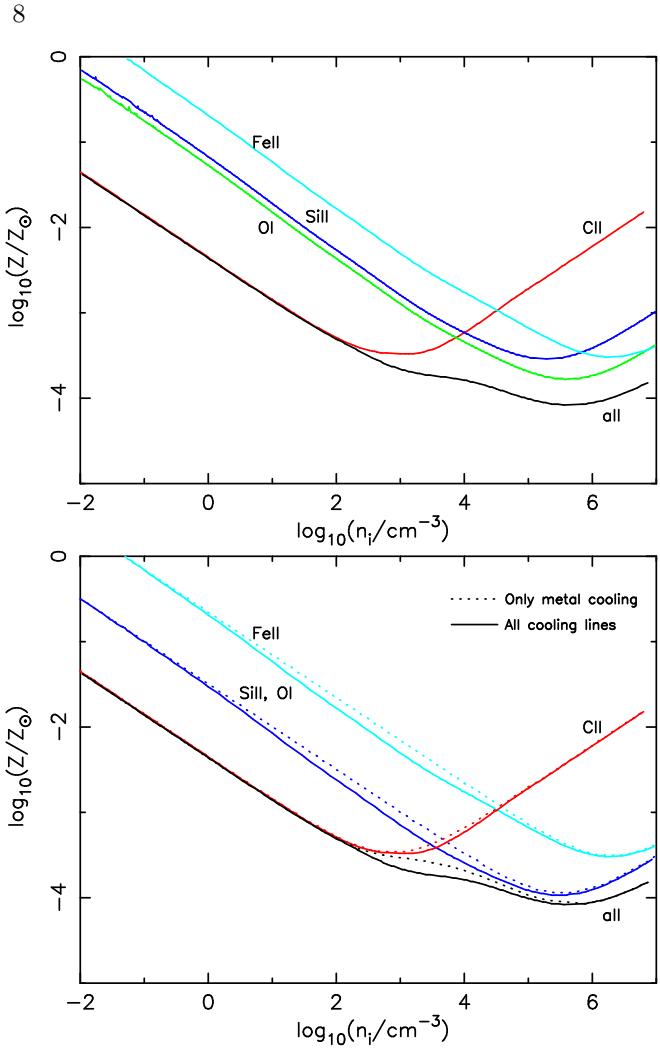

\psfig{width=8.7cm,angle=270,file=Fig5a.ps}
\psfig{width=8.7cm,angle=270,file=Fig5b.ps}
\caption{Top panel. Minimum critical metallicities for individual heavy
  elements vs.\ total number density at $T = 200$~K.  Curves 
  correspond to gas enriched only by C~II (dashed), Si~II (dash-dot), 
  O~I (dash-triple-dot), and Fe~II (dotted).  Solid curve shows all four 
  species together.  Bottom panel. Solid lines are simulations 
  that use a fragmentation criterion that equates adiabatic heating 
  to cooling by metals only, while dotted lines assume cooling from 
  metals plus \htwo\ (see \S~\ref{sec:modelm}).  Minima occur at 
  high densities, with log~$Z_{\rm crit} = -3.48$ (C~II), $-3.54$ (Si~II), 
  $-3.78$ (O~I), $-3.52$ (Fe~II), and $-4.08$ (all elements). } 
\label{fig:minimetallicity}
\end{figure}

The minimum critical metallicities are calculated from 
Figure \ref{fig:paracoolmcool} and shown in Figure \ref{fig:minimetallicity}. 
The bottom panel of Figure \ref{fig:minimetallicity} shows two sets 
of curves, to illustrate the slight differences in fragmentation 
criteria when one equates $\Gamma_{\rm ad}$ to cooling by metals-only 
(dashed lines) and to total cooling (solid lines). In both cases, one 
can see the transition from non-LTE to LTE of particular emission lines. 
The minimum value in each curve corresponds to the critical
density of the line (see Table \ref{tab:mabundances}). 
The curves turn around when collisional de-excitation 
of the lines starts to dominate for each metal. The volume cooling
rate then scales as $n$ rather than $n^2$. The cooling efficiency is 
reduced, and higher abundances of metals are needed to reach 
fragmentation.

Table \ref{tab:minimumcritical} shows the values of the minimum
critical metallicities for the fragmentation criterion, and for 
several values of gas density. We choose a range in densities, from
$n = 0.1$~cm$^{-3}$, just below the mean density of gas in virialized 
halos at $z = 20$, up to $n = 10^4$~cm$^{-3}$, characteristic of
dense cores. Within these halos, and particularly at their center, 
one expects to find clumps of cooler gas near the threshold of 
gravitational instability.    
These star-forming cores may have cooled or been compressed to 
higher values.  There may also be multiple clumps within the halo.
These inhomogeneities probably play an important role in how the
first star formation is initiated. They may also affect how the 
heavy elements are incorporated into surrounding gas, how those
gas clouds are taken to critical metallicity, and how subsequent
generations of stars are produced.

\begin{table}
\begin{center}
\caption{Minimum Critical Metallicities\tablenotemark{a} }
\label{tab:minimumcritical}

\begin{tabular}{cccccc}
\hline\hline \\
   $n_i$      & C~II & (O~I,~Si~II) & Fe~II & All Metals \\
(cm$^{-3}$) & log~$Z_{\rm crit}$ & log~$Z_{\rm crit}$ & 
           log~$Z_{\rm crit}$  &  log~$Z_{\rm crit}$    \\ 
\hline \\
 0.1        & $-1.77$  & $-1.00$ & $-0.16$ & $-1.86$\\
 1.0        & $-2.35$  & $-1.53$ & $-0.69$ & $-2.36$\\
 1.0(1)     & $-2.84$  & $-2.07$ & $-1.23$ & $-2.87$\\
 1.0(2)     & $-3.29$  & $-2.57$ & $-1.73$ & $-3.31$\\
 1.0(3)     & $-3.38$  & $-3.15$ & $-2.29$ & $-3.67$\\
 1.0(4)     & $-3.23$  & $-3.59$ & $-2.75$ & $-3.79$\\
 1.0(5)     & $-2.72$  & $-3.91$ & $-3.15$ & $-4.01$\\
 1.0(6)     & $-2.22$  & $-3.92$ & $-3.49$ & $-4.05$\\
\hline \\
\tablenotetext{a}{Minimum values of log~$Z_{\rm crit}$ needed 
   for fragmentation of gas with initial temperature $T = 200$~K 
   and number densities $n_i$. } 
\end{tabular}
\end{center}
\end{table}

\begin{figure}
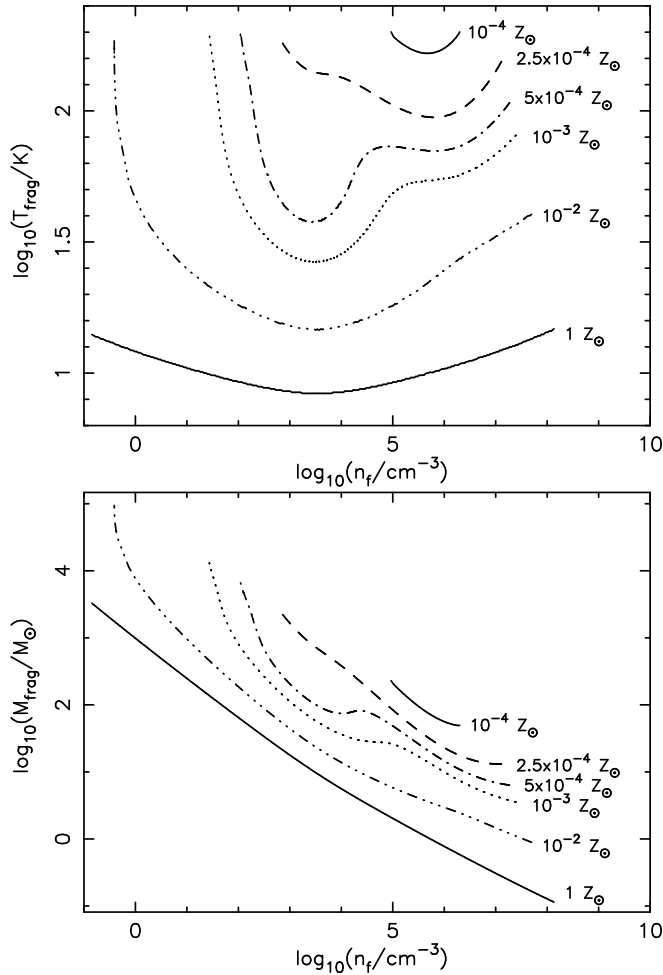

\psfig{width=8.7cm,angle=270,file=Fig6a.ps}
\psfig{width=8.7cm,angle=270,file=Fig6b.ps}
\caption{Top panel: fragmentation temperature as a function of the final 
  density, $n_f$, for gas with initial temperature $T=200$K.  Bottom panel:
  the Jeans mass at the point $(T_{\rm{frag}},n_{\rm{frag}})$~in units
  of solar masses.  All curves are labeled by the metallicity, ranging 
  from $Z=10^{-4}Z_\odot$ to $Z=Z_\odot$.} 
\label{fig:fragtempjmass}
\end{figure}

Figure \ref{fig:fragtempjmass} shows the final state of the gas at the 
instant of fragmentation. We fix the metallicity and evolve the cooling of 
the gas, for many initial densities, from $T = 200$~K until the fragmentation 
criterion is satisfied. The final temperature and Jeans mass of the fragments 
are plotted as a function of the final density of the gas. In both panels, 
we ran the simulation for five different metallicities, including 
all the metal lines. The curves correspond to metallicities ranging from 
$10^{-4}Z_\odot$ to $1~Z_{\odot}$.  Larger amounts of metals allow 
the gas to cool to very low temperatures, as expected. 
Figure \ref{fig:fragtempjmass} also illustrates the possibility of thermal phase 
transitions, resulting in a drop in the final temperature.  The temperature 
reaches a minimum at $n_f \approx 10^{3.5}$~cm$^{-3}$ for 
$-3.3 < \log~(Z/Z_{\odot}) < 0$, but rises to $n_f \approx 10^6$~cm$^{-3}$ 
for $Z \leq 10^{-3.3}~Z_{\odot}$.  This shift results from a
change in primary coolants, from [C~II] cooling ($n_{\rm cr} \approx 3000$ 
cm$^{-3}$) to fine-structure lines (\oi, \siii, \feii) with higher 
critical densities, $3 \times 10^5$ cm$^{-3}$ to $2 \times 10^6$~cm$^{-3}$.

\begin{figure}
\psfig{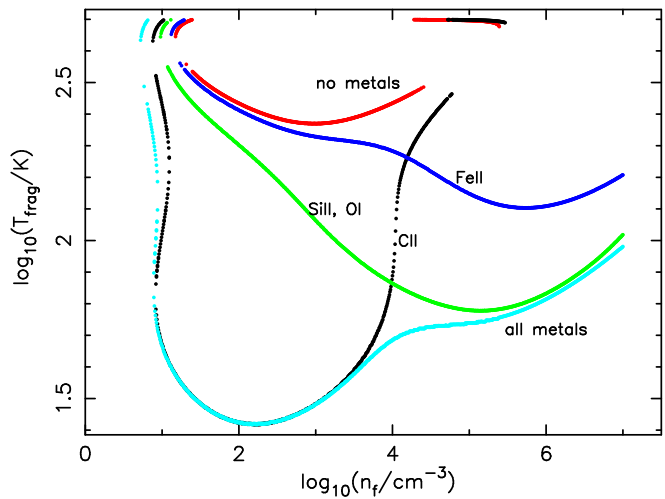}
\caption{Fragmentation temperature for a gas with no metals (red line),
  only Fe~II (dark blue line), only Si~II plus O~I (green line), 
  and only C~II (black line). The light blue includes all four metals. 
  In all cases, the metallicity is fixed at $Z=3\times10^{-4}Z_\odot$
  and the initial gas temperature is $T=200$~K.} 
\label{fig:fragtempmetals}
\end{figure}

Figure \ref{fig:fragtempmetals} shows the same final temperature as in
the top panel of Figure \ref{fig:fragtempjmass}, emphasizing the effects
of specific heavy elements.  As labeled, we show cases with no metal lines, 
only Fe~II, only Si~II plus O~I, only C~II, and all four elements in the 
cooling function. In each case, the metallicity is $Z=3\times10^{-4}Z_\odot$. 
At $n \approx 2\times10^4$~cm$^{-3}$, the cooling is driven mainly by
\oi\ lines.  This is because the C~II levels reach LTE, so its cooling 
scales as $n$ rather than $n^2$.   

\subsection{Multiple-Metallicity Simulations}
\label{sec:multiplez}

Empirical evidence from extremely metal poor (EMP) halo stars
\citep{Chr02} suggests that many progenitor stars, with a range of 
masses, contributed to the observed nucleosynthetic patterns.  
This point was stressed by Tumlinson, Venkatesan, \& Shull (2004) 
and Qian \& Wasserburg (2005). When exploring the concept
of ``critical metallicity", we need to understand the effects of
different mixtures of heavy elements, produced by massive stars
and SNe with a range of initial masses. 

We adopt the relative abundances 
of metals from studies of PISN yields. More specifically, we use 
the results from \cite{TVS04} who reviewed and critiqued the hypothesis 
that the primordial stars were VMS. They calculated the
theoretical yields from PISN for three progenitor mass ranges: 
150--170 $M_\odot$, 185--205 $M_\odot$, and 240--260 $M_\odot$
(see their Figure 1). We use these yields and run three sets of simulations,
in which we take the Fe abundance as a free parameter.  Depending on the 
range of masses and their yields, we calculate the metallicity of the 
other three elements (C, O, Si). Table \ref{tab:yields} gives the abundances 
of [C/Fe], [O/Fe], and [Si/Fe] and the mass ranges.
 
\begin{table}
\begin{center}
\caption{Metallicity yields from PISN\tablenotemark{a} } 
\label{tab:yields}

\begin{tabular}{ccccc}
\hline\hline\\
Mass Range & [C/Fe] & [O/Fe] & [Si/Fe] & [Si/O] \\[2pt] 
\hline 
$150-170~M_{\odot}$ &  1.4   &   1.9  &   2.4  & 0.5    \\ 
$185-205~M_{\odot}$ & $-0.2$ &   0.4  &   1.2  & 0.8    \\ 
$240-260~M_{\odot}$ & $-1.3$ & $-0.8$ & $-2.0$ & $-1.2$ \\
\hline
\tablenotetext{a}{Metallicity yields from PISN \citep{HW02} as 
   integrated by \cite{TVS04}. } 
\end{tabular}
\end{center}
\end{table}

\begin{figure}
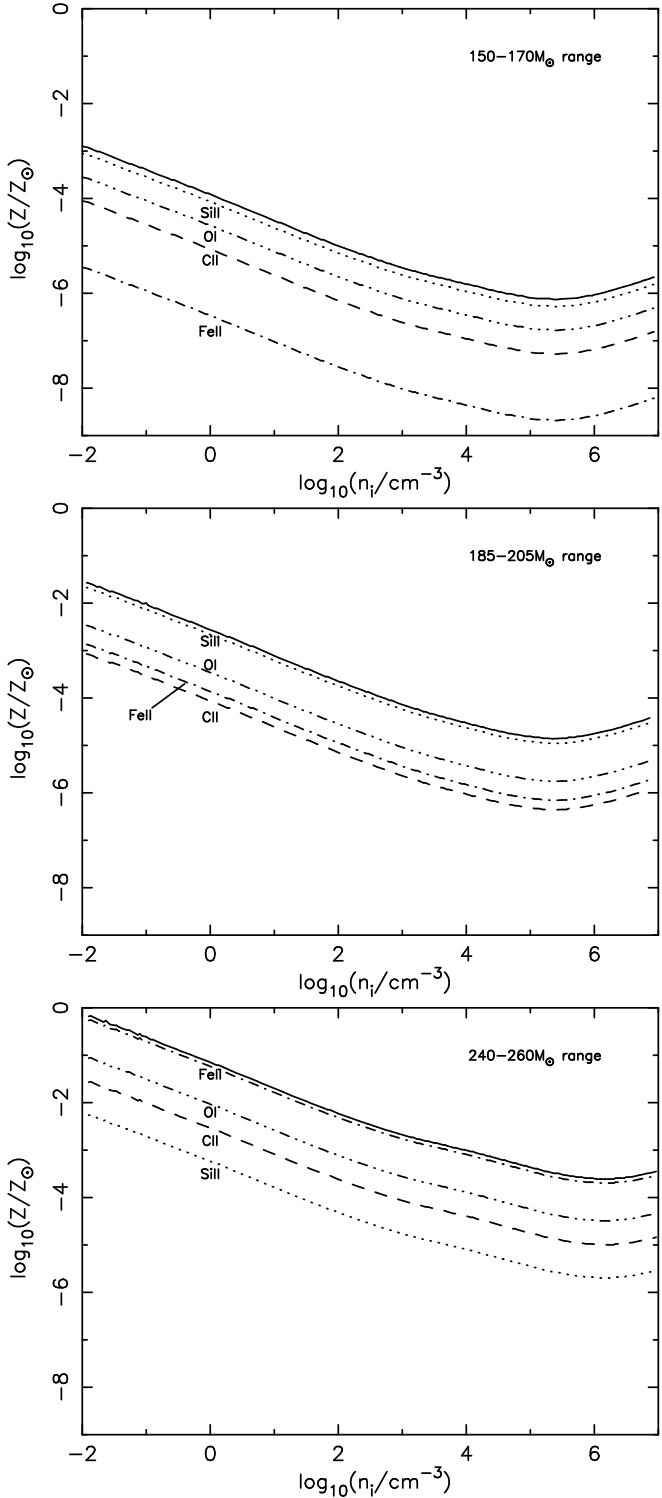

\psfig{width=8.7cm,angle=270,file=Fig8a.ps}
\psfig{width=8.7cm,angle=270,file=Fig8b.ps}
\psfig{width=8.7cm,angle=270,file=Fig8c.ps}
\caption{Minimum critical metallicity for fragmentation. Each plot
  corresponds to one of the three mass ranges of PISN yields. 
  The solid lines show the minimum critical metallicity,
  while the dotted and dashed lines are the components of the solid
  line, \cii, \oi, \siii, \feii. Top panel gives 150--170 M$_\odot$ 
  mass range, middle panel shows 185--205 M$_\odot$ range, and bottom
  panel shows 240--260 M$_\odot$ range. In the top panel, the dominant 
  cooling is from \siii, and in the bottom panel it is from \feii. 
  These shifts result in changes in $Z_{\rm crit}(n)$ relations,
  since Fe-rich gas has higher critical metallicity (see Fig.\ 5). } 
\label{fig:multiplem}
\end{figure}

Figure \ref{fig:multiplem} shows the minimum metallicity needed for 
fragmentation for three mass ranges of PISN. The solid line in all panels 
is the critical metallicity. For each mass range, $Z_{\rm crit}$ 
varies considerably, depending on the dominant elements synthesized. 
For a mass range 
150--170 $\Msun$, the main component of total $Z_{\rm crit}$ comes from 
Si~II, and the less important component is Fe~II. For the highest mass range, 
240--260 $\Msun$, the main metal cooling comes from Fe~II, while the least
important element is Si~II.  To a good approximation, C~II and O~I 
keep the same relative importance in all three mass ranges, 
with O~I contributing slightly more cooling than C~II.

\begin{figure}
\psfig{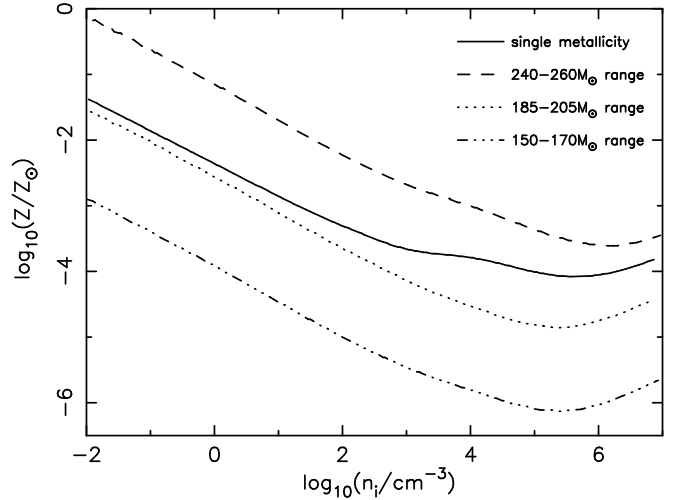}
\caption{This plot summarizes the two metallicity assumptions  
  used in this paper. Full line shows the total minimum critical
  metallicity from Figure \ref{fig:minimetallicity}. All four 
  elements appear in the cooling function with the same relative
  abundances. Dashed and dotted lines show the same results, but 
  with different relative abundances given by the yields of three 
  mass ranges of PISN. } 
\label{fig:simple-multiple}
\end{figure}

Figure \ref{fig:simple-multiple} summarizes all the results.  
The full line shows $Z_{\rm crit}$, considering a single metallicity, 
while the dashed and dotted lines show $Z_{\rm crit}$ from simulations 
with individual metallicities of C, O, Si, and Fe, given by the theoretical 
yields of PISN.

\subsection{Luminosities and Detection of Cooling Lines} 

In the preceding sections, we have explored the microphysics of
fine-structure cooling of primordial gas, as it makes a transition
from \htwo\ cooling to metal fine-structure line cooling.  An
obvious question is whether such cooling clumps are detectable
at high redshift.  How much mass must be cooling, and to what
metallicity must it be enriched, in order to rise above the
threshold of detectability by current and future telescopes?

For redshifts in the range $3 < z < 10$, the fine-structure 
lines under consideration will shift into the far-infrared (FIR)
and sub-millimeter regions of the spectrum.  On the ground, 
partial-transmission windows exist in the sub-mm, near 
350~$\mu$m (333--380~$\mu$m), 450~$\mu$m (420--500 $\mu$m), and 
850~$\mu$m (830--910~$\mu$m).  These FIR windows will be observable 
on the ground with ALMA.  In space, there are several planned
missions that offer a chance for more sensitive FIR observations
with mirrors cooled to 5--10~K. The SPICA mission is a proposed 
Japanese infrared astronomical satellite \citep{Matsumoto03} 
designed to explore the universe with a cooled (4.5~K) telescope 
of 3.5m aperture.  A spectrometer aboard SPICA \citep{Bradford03}  
could reach $5\sigma$ flux limits of $10^{-21}$ W~m$^{-2}$.  
An even more ambitious mission is SAFIR ({\it Single Aperture Far 
Infrared Telescope}), a 10m-class, cold (4--10~K) space telescope 
with hopes of achieving broad-band FIR sensitivity to emission-line 
fluxes of $10^{-22}$ W~m$^{-2}$ \citep{Lester04,Benford04}. 
     
To produce detectable FIR line emission, the cooling gas clouds 
must have baryon masses of at least $M_b = (10^8~M_{\odot})M_8$ 
at metallicities $Z \geq 0.01~Z_{\odot}$.  If the clouds have high
densities,  $n_H \geq n_{\rm cr}$, the cooling rate is proportional
to mass, and we can estimate the line luminosities from the LTE cooling 
rates, $L_i = {\cal L}_{\rm LTE}^{(i)}(T) (M_b/\mu)$.  Using the
results of \S~2.4 for $T \approx 200$~K, we estimate that the
metal fine structure lines have luminosities:
\begin{eqnarray}
  L_{\rm CII} &=& (3.7 \times 10^{39}~{\rm erg~s}^{-1})
            M_8 (Z_{\rm C}/0.01~Z_{\odot}) \\
  L_{\rm SiII} &=& (8.1 \times 10^{40}~{\rm erg~s}^{-1}) 
            M_8 (Z_{\rm Si}/0.01~Z_{\odot}) \\
  L_{\rm OI} &=& (2.0 \times 10^{41}~{\rm erg~s}^{-1}) 
            M_8 (Z_{\rm O}/0.01~Z_{\odot}) \\
  L_{\rm FeII} &=& (2.2 \times 10^{41}~{\rm erg~s}^{-1}) 
            M_8 (Z_{\rm Fe}/0.01~Z_{\odot}) \; .  
\end{eqnarray}  
A similar calculation can be made for the first two rotational
lines of \htwo, at 28.22~$\mu$m and 17.03~$\mu$m, assuming LTE 
populations for the upper rotational levels, $J = 2$ and 3.
This calculation differs somewhat from the numerical calculations,
in which we fix the ortho/para ratio at a constant value 3:1. 
For the 2-level analytic model, 
\begin{eqnarray}
  L_{28\mu m} = (1.4 \times 10^{37}~{\rm erg~s}^{-1})
            M_8 \, (f_{\rm H2}/0.001)    \\ 
 L_{17\mu m} = (1.2 \times 10^{38}~{\rm erg~s}^{-1})
            M_8 \, (f_{\rm H2}/0.001)   
\end{eqnarray}

\begin{figure}
\psfig{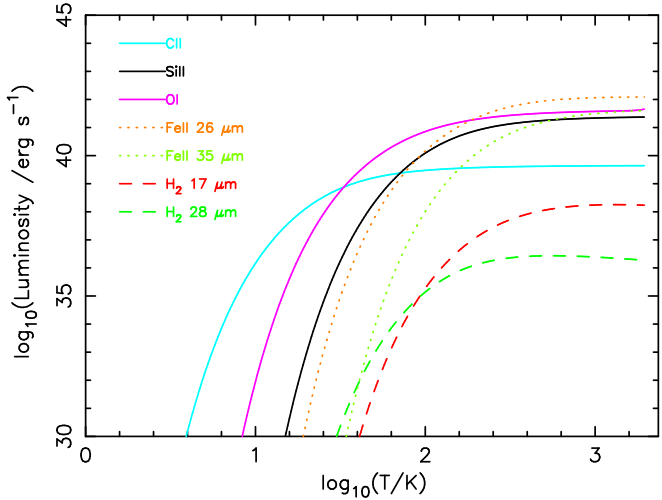}
\caption{Individual line luminosities for $10^8~M_{\odot}$ of
gas at various temperatures, for fine-structure lines of 
\cii\ (158 $\mu$m), \oi\ (63~$\mu$m), \siii\ (34.8~$\mu$m), 
\feii\ (25.99~$\mu$m, 35.35~$\mu$m), and the lowest rotational 
lines of \htwo\ at 28.22~$\mu$m and 17.03~$\mu$m.  These models 
assume maximally efficient cooling, with high-density LTE 
populations,  a single metallicity $Z = 0.01~Z_{\odot}$, and 
molecular fraction $f_{\rm H2} = 2 \times 10^{-4}$. } 
\label{fig:line-luminosities}
\end{figure}

Figure \ref{fig:line-luminosities} shows the individual
line luminosities from a numerical model with $10^8~M_{\odot}$
of cooling gas at $10^{-2}~Z_{\odot}$ and residual
molecular fraction $f_{\rm H2} = 2 \times 10^{-4}$ at 
temperatures ranging from 10--1000~K. Note that the
metal fine-structure lines of \oi, \siii, and \feii\ are
considerably stronger than the \htwo\ rotational lines.    
Even for metallicities ($Z_{\rm O} \approx 0.001$)
and molecular fraction ($f_{\rm H2} \approx 0.001$), the
\oi\ fine-structure line is 200 times stronger than
the 17.03~$\mu$m line of \htwo\ and 1000 times stronger
than the 28.22~$\mu$m line.

At redshifts $z \approx 4$, the source luminosity distances are
$d_L \approx (c/H_0)[z (1+z/4)] \approx (10^{29}~{\rm cm})h_{70}^{-1}$. 
Luminosities of $(10^{41}~{\rm erg~s}^{-1})L_{41}$ 
therefore correspond to line fluxes 
$(8 \times 10^{-22}~{\rm W~m}^{-2})~L_{41}$, where we have converted 
from cgs units to SI units of line sensitivity (W~m$^{-2}$) used by 
the SAFIR studies. At $z \approx 10$, $d_L$ is 4 times larger, so
these fluxes would drop by another factor of 16.   

For a given metallicity, $Z$, the strongest lines from metal-enriched
clouds are probably \oi\ 63.18~$\mu$m and \feii\ 25.99~$\mu$m.  Lines
of \siii\ (34.8~$\mu$m) and \cii\ (158~$\mu$m) are ten times weaker,
at the same metallicity.  However, as we have noted above, it is 
possible that O, Si, and Fe metallicities could be enhanced
in regions contaminated by products of massive stars. For
$10^8~M_{\odot}$ of cooling gas at 1\% solar metallicity and
$T \approx 200$~K, these lines should have luminosities  
$\sim 2 \times 10^{41}$ erg~s$^{-1}$ and fluxes of 
$(1-2) \times 10^{-21}$~W~m$^{-2}$ at $z = 4-5$.  Ground-based searches 
for these lines in the $350~\mu$m band would probe redshift intervals 
$z_{\rm OI} = 4.3-5.0$ and $z_{\rm FeII} = 11.8-13.6$.  Searches at
other redshifts open up in the space FIR.

\section{Discussion and Conclusions}
\label{sec:discuss}

In this paper, we have examined the thermal evolution of a primordial 
gas that has been enriched with trace amounts of metals. We have 
assumed that these metals have managed to leave their parent halos
\citep{TVS04}, where the first stars have exploded as supernovae.
Therefore, a maximum level of metallicity is reached in the IGM. The 
purpose of this paper was to calculate the fraction of metals 
necessary to produce continuum fragmentation and to examine the
properties of the gas when this occurs.

The results of this paper have been divided in two parts, in the first
part (\S~\ref{sec:singlez}), we assumed a single metallicity, with 
C, O, Si, and Fe appearing in relative solar abundances. We used this
metallicity as a exploratory parameter to find the instant at which the 
gas fragments. The second part of our study considered individual 
metallicities (\S~\ref{sec:multiplez}). This is a more natural approach, 
in which the fractional abundance for each metal comes from theoretical
PISN yields, averaged over various mass ranges, calculated by
\cite{TVS04}. In this case, we used Fe as a exploratory parameter and
the amount of metallicity for each of the other three metals considered 
in this paper depends on the mass range of the PISN.

We have neglected in this work any kind of radiative feedback (UV or X-ray),
more specifically photo-ionization heating, which raises the temperature of 
the gas and therefore shutting off star-formation (the temperature
raises over $>10^4$~K). Another neglected heating process is 
cosmic rays, which will be examined in a later paper.
To compare our results with work done by previous authors, we 
looked for the point in clump evolution when the cooling rate 
($\Lambda_{\rm{CII}, \rm{OI}, \rm{SiII}, \rm{FeII}}$) equals the 
adiabatic compressional heating rate.  This is the definition used by 
\cite{BL03} for \cii\ and \oi\ cooling. We have also introduced 
fine-structure cooling by \siii\ and \feii, and we explored the
slight differences when one equates the total cooling rate (including 
molecular cooling) to the compressional heating rate. 

Our results at high densities agree well with \cite{BL03}. However, 
the required level of metallicity for a gas to fragment at high redshift 
depends on whether the second-generation star formation is triggered 
from clumps or from low density gas within the halo.  As can be
seen from Figures 4 and 8, the critical metallicity is considerably
higher ($\sim1$\% solar) for the mean gas density,
$n_{\rm H} \sim 0.30$~cm$^{-3}$, typical of $10^6~M_{\odot}$ virialized 
halos at $z \approx 20$.  These halos may have denser clumps, into
which the first heavy elements are incorporated, but the details of 
these processes are complicated (Tumlinson, Venkatesan, \& Shull 
2004).  Numerical simulations are probably required to fully understand 
the metal transport and fraction of metals that escape into the 
lower-density IGM.   

At densities $n > n_{\rm cr}$ for the important fine-structure
coolants, the fragmentation criterion shifts $Z_{\rm crit}$ to
higher values.  Cooling is most efficient at or near the densities,
$n \approx n_{\rm cr}$, at which the fine-structure levels reach LTE.  
For this density environment, the minimum critical metallicities,
$Z_{\rm crit}$, are [C/H]$_{\rm crit} \approx -3.48$, 
[O/H]$_{\rm crit} \approx -3.78$, [Si/H]$_{\rm crit} = -3.54$, and 
[Fe/H]$_{\rm crit} \approx -3.52$.  When all metal transitions are 
included, we find a minimum $Z_{\rm crit} = 10^{-4.08}~Z_{\odot}$  
and $M_J \approx 117~M_{\odot}$, with a temperature 
$T_{\rm crit} = 199$~K. In this case, all the metals appear in the
cooling function with the same fractional abundance, but these results
are highly dependent on the amount of mixing in the gas, which
may determine the relative abundances of each metal.

An important parameter to consider in more detail is the initial
abundance of residual molecular hydrogen. In all the simulations
presented in this paper, the initial fractional abundance was taken 
as the residual \htwo\ fraction, 
$f_{\rm H2,res} = n_{\rm H2,res}/n_{\rm tot} = 2.0 \times 10^{-4}$. 
We also ran the single-metallicity simulations with 
$f_{\rm H2,res} = 1.1\times10^{-6}$~\citep{GaP98}.  The differences 
between simulations run with low-\htwo\ and high-\htwo\ 
residual abundances are small at low densities.  The maximum
difference in $Z_{\rm crit}$ is $\sim50$\% at 
$n_f \approx 2000$ cm$^{-3}$. 

For excitation from the ground vibrational state of \htwo, the most
important rotational cooling lines are at $28.22~\mu$m 
($J = 2 \rightarrow 0$) and $17.03~\mu$m ($J = 3 \rightarrow 1$),
labeled (0-0) S(0) and (0-0) S(1), respectively.  Because of
their low radiative transition rates, these \htwo\ emission lines
are quite weak; the metal fine-structure lines are expected to
be considerably stronger, even at $T = 200$~K, once the metallicity
rises above $10^{-4}~Z_{\odot}$.   

Using individual metallicities for C, O, Si, and Fe, we found an 
important distinction from the single-metallicity case. These results 
suggest that, for massive stars enhanced in $\alpha$-process 
elements (O, Si and Fe), the critical metallicity for fragmentation is 
much higher than previous works have suggested. Higher metallicities
are also required if the metal-enriched gas has lower densities
than $n_{\rm cr}$. 

Redshifted fine-structure lines from metal-enriched primordial gas
at 200~K could be observable from high-redshift halos.  In addition 
to the weak rotational lines of \htwo\ (28~$\mu$m and 17~$\mu$m) and 
\cii\ 157.74~$\mu$m, the most important coolants are likely to be 
\oi\ 63.18~$\mu$m, \siii\ 34.8~$\mu$m, and \feii\ (25.99~$\mu$m
and 35.35~$\mu$m). With 
redshifting, these lines would appear in the FIR sub-mm, and even the 
mm-band, if $(1+z) \approx 10-20$.      

\section*{Acknowledgements}
We are grateful to Aparna Venkatesan, Jason Tumlinson, Jason Glenn,
and Dan Lester for useful discussions regarding this project, and 
Phil Maloney for help with the numerical techniques.  This research
was supported by astrophysics theory grants from NASA (NAG5-7262)
and NSF (AST02-06042).

\bibliographystyle{apj} 
\bibliography{zcrit}


\newpage

\renewcommand{\theequation}{A.\arabic{equation}}\setcounter{equation}{0}

\begin{appendix}

\section{Primordial Chemical Reaction Network}

Table \ref{tab:hmtab} shows the main chemical reaction between elements
in the primordial gas used in our simulations. The gas is composed by
\htwo,  \h,  \hplus, \htwoplus, \hminus,  \he,  \heplus, \heplusplus, 
and  \e. The reaction rates were compiled from the following
authors: \cite{HTL96}, \cite{GaP98}, \cite{FuC00}, \cite{SLD98} and
\cite{AAZN97}. The minimal model shown in the table was defined
by \cite{HST02}.

\begin{table}
\centering
\caption{Chemical Reactions and Rates\tablenotemark{a}}   

\label{tab:hmtab}
\begin{tabular}{lllr}
\hline\hline \\ 
\emph{ } & Reaction& Rate (cm$^3$~s$^{-1}$) & Reference \\[2pt]
\hline\\[2pt]

 1 & $\h + \e \longrightarrow \hplus + 2\e$
   & $5.9\times10^{-11}\,T_0^{0.5}(1+T_5^{0.5})^{-1}\exp(-1.58/T_5)$
   & HTL \\[2pt]

 2 & $\hplus + \e \longrightarrow \h + \gamma$ 
   & $3.3\times10^{-10}\,T_0^{-0.7}(1+T_6^{0.7})^{-1}$
   & HTL \\[2pt]

 3 & $\h + \e \longrightarrow  \hminus + \gamma$ 
   & $1.4\times 10^{-18}\,T_0^{0.93}\exp(-T_4/1.62)$
   & GP \\[2pt]


 4 & $\hminus + \h \longrightarrow \htwo + \e$
   & $1.3\times10^{-9}$
   & FC \\[2pt]

 5 & $\hminus +\hplus \longrightarrow 2\h $
   & $4.0\times 10^{-6}\ T_0^{-0.5}$
   & FC\\[2pt]

 6 & $\h + \hplus \longrightarrow \htwoplus + \gamma$
   & $2.1\times10^{-23}\,T_0^{1.8}\exp(-2/T_1)$
   & SLD\\[2pt]


 7 & $\htwoplus + \h \longrightarrow \htwo + \hplus$
   & $6.4\times 10^{-10}$
   & GP \\[2pt]

 8 & $\htwoplus + \e \longrightarrow 2\h $
   & $1.2\times 10^{-7}\ T_0^{-0.4}$
   & SLD \\[2pt]

 9 & $\htwo + \hplus \longrightarrow \htwoplus +\h$
   & $\min \big[3.0\times 10^{-10}\exp(-2.11/T_4)$,
     $1.5\times 10^{-10}$\\
   &
   & $\exp(-1.40/T_4)\big]$
   & GP \\[2pt]

10 & $\htwo + \h \longrightarrow 3\h$
   & $7.1\times 10^{-19}\,T_0^{2.01}(1+2.13\,T_5)^{-3.51}$\\
   &
   & $\exp\left(-5.18/T_4\right)$
   & AAZN \\[2PT]

11 & $\htwo + \e \longrightarrow 2\h + \e$
   & $4.4\times 10^{-10}\,T_0^{0.35}\exp(-1.02/T_5)$
   & GP \\[2pt]

12 & $\he + \e \longrightarrow \heplus + 2\e$
   & $2.4\times 10^{-11}\,T_0^{0.5}\,(1+T_5^{0.5})^{-1}\exp(-2.85/T_5)$ 
   & HTL \\[2PT]

13 & $\heplus + \e \longrightarrow \he + \gamma$
   & $ 1.5\times10^{-10}\,T_0^{-0.64}
     +1.9\times10^{-3}\,T_0^{-1.5}$\\
   &
   & $\exp(-5.64/T_5)\,(0.3+\exp(9.40/T_4))$ 
   & AAZN \\[2PT]

14 & $\heplus + \e \longrightarrow \heplusplus + 2\e$
   & $5.7\times10^{-12}\,T_0^{0.5}\,(1+T_5^{0.5})^{-1}\exp(-6.32/T_5)$ 
   & HTL \\[2PT]

15 & $\heplusplus + \e \longrightarrow \heplus + \gamma$
   & $1.3\times10^{-9}\,T_0^{-0.7}\,(1+T_6^{0.7})^{-1}$
   & HTL \\[2PT]  

\hline \\

\tablenotetext{a}
{This table summarizes the important reactions needed to calculate 
accurately the abundances of \htwo\ and \e. References are: HTL
Haiman, Thoul \& Loeb (1996); GP Galli \& Palla (1998); FC Fuller \&
Couchman (2000); SLD Stancil, Lepp \& Dalgarno (1998); AAZN Abel,
Anninos, Zhang \& Norman (1997). } 

\end{tabular}
\end{table}

\end{appendix}

\end{document}